\documentclass[%
 reprint,
 superscriptaddress,
 amsmath,amssymb,
 aps,
]{revtex4-1}

\usepackage{graphicx}
\usepackage{dcolumn}
\usepackage{bm}
\usepackage{xcolor}

\begin{document}


\title{The European Muon Collaboration effect from short-range
correlated nucleons in a nucleon swelling model}  

\author{Na-Na Ma}
\email{mann@lzu.edu.cn}
\affiliation{School of Nuclear Science and Technology, Lanzhou University, Lanzhou 730000, China}

\author{Tao-Feng Wang}
\email{tfwang@buaa.edu.cn}
\affiliation{School of Physics, Beihang University, Beijing 100191, China}
\affiliation{China Institute of Atomic Energy, Beijing 102413, China}

\author{Rong Wang}
\email{Corresponding author: rwang@impcas.ac.cn}
\affiliation{Institute of Modern Physics, Chinese Academy of Sciences, Lanzhou 730000, China}
\affiliation{School of Nuclear Science and Technology, University of Chinese Academy of Sciences, Beijing 100049, China}


\date{\today}

\begin{abstract}
The relation between the nuclear EMC effect and the nucleon-nucleon short-range
correlation is a hot topic in high-energy nuclear physics,
ever since a peculiar linear correlation between these two phenomena discovered.
In this paper, the contribution to the nuclear EMC effect arising from the
short-range correlated nucleons is examined in a nucleon-swelling model.
We find that the structure modifications of the N-N SRC nucleons reproduce
more or less the measured EMC ratios of light nuclei,
while they are not enough to explain the measured EMC ratios of heavy nuclei.
We speculate that the hypothesis of causal connection between SRC and the EMC effect is
not exact, or the universality of the inner structure of SRC nucleon is violated
noticeably from light to heavy nuclei, or there are other origins for the EMC effect.
\end{abstract}

\pacs{21.60.Gx, 24.85.+p, 13.60.Hb}
\maketitle


\section{Introduction}
\label{sec:intro}

The nuclear EMC effect observed in lepton-nucleus deep
inelastic scattering (DIS) \cite{EuropeanMuon:1983wih,EuropeanMuon:1988lbf,Bodek:1983qn} proves that the quark degrees
of freedom inside nucleon are influenced by the surrounding
nucleons (cold nuclear medium). This phenomenon implies
that the nuclear force between nucleons is emergent fundamentally from
the strong interaction between the quarks inside different nucleons.
Before the experiment of EMC collaboration, the quark degrees of freedom are
thought to be frozen and confined in the nucleon, and the nuclear
force at the scale around nuclear binding energy can not influence
the nucleon inner structure to a sizeable extent.
It attracted a lot of interests soon after the discovery
and it is still an interesting puzzle in high-energy
nuclear physics through decades \cite{Arneodo:1992wf,Geesaman:1995yd,Norton:2003cb,Hen:2013oha,Malace:2014uea}.
Understanding of the mechanism of the EMC effect from
quantum chromodynamics (QCD) remains quite challenging \cite{Detmold:2020snb,Winter:2017bfs}.

The nucleon-nucleon short-range correlation (N-N SRC) is one microscopic
and quite unusual structure inside an atomic nucleus \cite{Hen:2016kwk,Arrington:2011xs,Frankfurt:1988nt,Fomin:2017ydn,Arrington:2022sov}.
Different from the mean-field description of the nuclear interaction
and the single-nucleon motion given by the nuclear shell model,
the N-N SRC shows one kind of special close-proximity structure
of the nucleon-nucleon distance about or even smaller than 1 fm \cite{Hen:2016kwk,Fomin:2017ydn}.
In the N-N SRC pair, the nucleon-nucleon interaction can reach the
repulsive core of the nuclear force. Therefore the nucleon struck out
from N-N SRC could have the momentum way higher than the nuclear Fermi momentum.
Thanks to the clean probe of high-energy electron,
the N-N SRC is observed in the inclusive and exclusive processes,
identified with the high nucleon momentum
and the angle correlation between the high-momentum nucleon partners
\cite{Frankfurt:1993sp,CLAS:2005ola,Fomin:2011ng,Subedi:2008zz,CLAS:2018yvt,Hen:2014nza,CLAS:2018xvc,CLAS:2018qpc,CLAS:2020mom,CLAS:2019vsb}.
Though the short-range correlated nucleons interact extensively and strongly,
they are the minorities in the nucleus compared to the mean-field nucleons.
In heavy nuclei, only about 20\% of the nucleons are in the N-N SRC configuration
\cite{CLAS:2005ola}.

There is no doubt that the nucleons in close-proximity interact
with each other strongly. Their inner structures therefore
can be greatly modified. Naively, the N-N SRC is thus thought to be
an important source of the EMC effect.
Actually, with the finding of a linear correlation between
the magnitude of EMC effect and the relative number of N-N SRC pairs \cite{Weinstein:2010rt,Hen:2012fm},
more and more physicists guess that the strong modification of SRC
nucleons is the primary origin of the EMC effect.
Theoretically, the linear correlation between the EMC effect
and the N-N SRC is explained with the scale separation phenomenon \cite{Chen:2016bde}.
Experimentally, the CLAS collaboration tested the SRC-driven model
for the nuclear EMC effect, with the simultaneous measurements
of DIS and quasi-elastic inclusive process on the deuteron
and some heavier nuclei. They extracted the modification function
of the structure function of SRC nucleon and found that
this modification function is more or less universal for different nuclei \cite{CLAS:2019vsb}.
A latter study examined the SRC-driven EMC effect and the universal
modification function in both the local-density and the high-virtuality
pictures, and a truly universal modification function of SRC was found \cite{Arrington:2019wky}.
It is proposed that the EMC effect is not the traditional
static modification on all the independent nucleons
but a strong dynamical effect from two strongly interacting nucleons
fluctuating into a temporary high-local-density SRC pair.

However, different people have different opinions in explaining
the correlation between the EMC effect and the N-N SRC.
The relationship of these two phenomena is examined recently
in details with a convolution model which incorporates the nuclear binding
and the nucleon off-shell effects \cite{Wang:2020uhj}.
They argue that their analysis does not support the hypothesis
that there is a causal connection between SRC nucleons and the EMC effect.
The EMC effect of the low-momentum nucleon and
the high-momentum nucleon are studied separately.
They find that the Fermi motion effect overwhelms the off-shell
effect for the SRC nucleons with various models for the off-shell correction.
Thus they conclude that the SRC nucleons do not give a dominant part of
the observed EMC effect \cite{Wang:2020uhj}, compared with the mean-field nucleons.
In our previous paper \cite{Wang:2022kwg}, we also get the similar conclusion that
only the modifications on the N-N SRC nucleons are not enough to
reproduce the measured EMC effect, with our current knowledge
about the number of SRC pairs in the nuclei \cite{CLAS:2005ola,CLAS:2018yvt,Wang:2020egq}.
In our previous analysis, the $x$-rescaling model is applied
for the off-shellness correction of the SRC nucleon,
and the effective mass of SRC nucleon is taken from a recent analysis \cite{Wang:2020egq}.

In this paper, the hypothesis that the nuclear EMC effect comes entirely form
the N-N SRC pairs is examined further at a more fundamental level.
The conventional nuclear models usually take into account the reduced
nucleon mass in medium or the nucleon virtuality for the EMC effect,
leading to the $x$-rescaling models
\cite{GarciaCanal:1984eh,Staszel:1983qx,Akulinichev:1985xq,Frankfurt:1985ui,Jung:1988jw,CiofidegliAtti:1999kp}
and the off-shellness corrections \cite{Dunne:1985ks,Gross:1991pi,Kulagin:1994fz,Kulagin:2004ie,Kulagin:2014vsa}.
Since the EMC effect is measured in the DIS process,
it should be explained at the quark level instead of the nucleon level.
The QCD-inspired models in explaining the EMC effect usually require
an increase of the quark confinement or a simple picture of nucleon swelling.
As the nucleons in SRC pair are so close to each other and forms
into a high-local density cluster, the quarks inside could thus be de-confined.
In the hadron bag picture, we can imagine the two nucleon bags
merge into a big di-nucleon bag.
If the quarks can move freely from one nucleon to the other
in the SRC pair, then the confinement space of the quark could be
enlarged by as big as twice.
Within the nucleon swelling model, the quark distributions inside
the SRC nucleon can be calculated quantitatively \cite{Wang:2018wfz,Wang:2016mzo}.
Hence the contribution of the SRC nucleons
to the EMC effect can be evaluated.

The organization of the paper is as following.
The hypothesis that the nuclear EMC effect arises dominantly
from the N-N SRC pairs and the related formula are given in Sec. \ref{sec:EMC-and-SRC}.
The nucleon swelling model for calculating the structure function
of the SRC nucleon is discussed in Sec. \ref{sec:swelling-effect}.
The results of the SRC driven model for the EMC effect are shown in Sec. \ref{sec:results}.
A brief summary of the analysis is given in Sec. \ref{sec:summary}.

\section{Nuclear EMC effect from N-N SRC}
\label{sec:EMC-and-SRC}

A haunting question we try to answer in this work is whether
the N-N SRC is wholly responsible for the nuclear EMC effect.
Therefore we employ the so-called ``SRC-driven model'' for the EMC effect.
That means: the inner structure of short-range correlated nucleons are
substantially modified while the inner structure of the nucleons
in the mean field are nearly unmodified.
The N-N SRC is the only (or dominant) source of the EMC effect.
The long-range nuclear interaction has no influence on the
short-distance structure in the nucleon.

Many experiments have revealed that the majority of
N-N SRC pairs are the proton-neutron correlated pairs
\cite{Subedi:2008zz,CLAS:2018yvt,Hen:2014nza,CLAS:2018xvc}.
This isophobic property is actually consistent with the theoretical calculations
based on the assumption that the medium-range tensor force is primarily responsible
for the formation of N-N SRC pairs \cite{CLAS:2020mom,Schiavilla:2006xx,Alvioli:2007zz,Neff:2015xda}.
In this paper, we study the model which assumes that
the N-N SRC is the primary source of the EMC effect.
For the simplicity of model calculations,
we ignore the p-p and n-n SRC pairs,
since they together are surely the minorities ($\lesssim$10\%) compared to the p-n SRC pairs.
Thus the per-nucleon nuclear structure function is given by,
\begin{equation}
\begin{split}
F_2^{\rm A}=&\left[n^{\rm A}_{\rm SRC}F_2^{\rm p~in~SRC}+n^{\rm A}_{\rm SRC}F_2^{\rm n~in~SRC}\right. \\
&\left.+(Z-n^{\rm A}_{\rm SRC})F_2^{\rm p}+(A-Z-n^{\rm A}_{\rm SRC})F_2^{\rm n} \right] /A,
\end{split}
\label{eq:SRC-driven-EMC-model}
\end{equation}
in which $n^{\rm A}_{\rm SRC}$ is the number of p-n SRC pairs in nucleus $A$,
$F_2^{\rm p~in~SRC}$ and $F_2^{\rm n~in~SRC}$ are the modified nucleon structure functions
in the SRC pair, and $F_2^{\rm p}$ and $F_2^{\rm n}$ are the free nucleon structure functions.
$Z$, $N$ and $A$ are respectively the proton number, neutron number
and the mass number to define a particular nucleus.
Note that the universality of p-n SRC pair
in different nuclei is assumed for Eq. (\ref{eq:SRC-driven-EMC-model}).

The N-N SRC is a compact and short-time lived state
from the fluctuations of the many-body dynamics of nuclear force.
The formations and dissociations of N-N SRC pairs keep on going inside the nucleus.
Thus in Eq. (\ref{eq:SRC-driven-EMC-model}),
the number of SRC pairs $n_{\rm SRC}$ should be
viewed as a mean value in the measurements.
Take the deuteron for an example, the mean number of p-n SRC pairs
in the deuteron is less than one ($n^{\rm d}_{\rm SRC}\ll 1$),
for the N-N SRC configuration happens very occasionally.

For the SRC-driven model, the number of SRC pairs
in a nucleus ($A$) is an indispensable parameter.
In experiment, the relative number of N-N SRC pairs is characterized
by the SRC scaling ratio $a_2$ in the region $1.4\lesssim x_B \lesssim 1.9$.
Then the number of SRC pairs in nucleus $A$ $n^{\rm A}_{\rm SRC}$ is computed
with the measured $a_2$ and the number of SRC pairs $n^{\rm d}_{\rm SRC}$ in deuteron,
which is written as,
\begin{equation}
n^{\rm A}_{\rm SRC}=[A\times a_2(A)\times n_{\rm SRC}^{\rm d}]/2.
\label{eq:SRCPairNumber}
\end{equation}
The SRC scaling ratio $a_2$ are measured using the high-energy electron inclusive
scattering process off the nuclear targets \cite{CLAS:2005ola,Fomin:2011ng,CLAS:2019vsb}.
The number of SRC pairs in deuteron has already been determined
in our previous analysis \cite{Wang:2020egq}.
Table \ref{tab:a2-values} lists the values of $a_2$ of some nuclei,
measured by CLAS collaboration \cite{CLAS:2005ola,CLAS:2019vsb}
and JLab Hall C collaboration \cite{Fomin:2011ng},
and also the averaged values that used in this analysis.
Note that in Eq. (\ref{eq:SRCPairNumber}), the small effect of
the pair motions \cite{CLAS:2005ola,Fomin:2011ng} is not considered. Contrary to the SRC universality,
the pair center-of-mass (c.m.) motion effect is nuclear-dependent.

\begin{table}[h]
    \caption{The experimental data of SRC scaling factor $a_2$ from
             CLAS and JLab Hall C collaborations,
             and the resulting average values, for various nuclei. }
        \renewcommand\arraystretch{1.25}
		\begin{tabular}{ccccc}
        \hline\hline
            Nucleus  & CLAS06\cite{CLAS:2005ola} & CLAS19\cite{CLAS:2019vsb} & Hall C\cite{Fomin:2011ng} & Average   \\
        \hline
            $^3$He   &  1.97$\pm$0.10  &                & 2.13$\pm$0.04 & 2.11$\pm$0.04    \\
            $^4$He   &  3.80$\pm$0.34  &                & 3.60$\pm$0.10 & 3.62$\pm$0.10    \\
            $^9$Be   &                 &                & 3.91$\pm$0.12 & 3.91$\pm$0.12    \\
          $^{12}$C   &  4.75$\pm$0.41  & 4.49$\pm$0.17  & 4.75$\pm$0.16 & 4.64$\pm$0.11    \\
         $^{27}$Al   &                 & 4.83$\pm$0.18  &               & 4.83$\pm$0.18    \\
         $^{56}$Fe   &  5.58$\pm$0.45  & 4.80$\pm$0.22  &               & 4.95$\pm$0.20    \\
        $^{208}$Pb   &                 & 4.84$\pm$0.20  &               & 4.84$\pm$0.20    \\
        \hline\hline
		\end{tabular}
    \label{tab:a2-values}
\end{table}

The other important input for the model of SRC-induced EMC effect are
the modified structure function of SRC nucleon.
The structure function at intermediate $x_B$ is closely related
to the valence quark distributions.
A model derived from the expansion of quark confinement is employed to estimate
the quark distributions and the structure function of the SRC nucleon.
We discuss such model in details in the following section.

\section{Swelling effect for SRC nucleons}
\label{sec:swelling-effect}

How we compute the structure functions of the free nucleon
and the SRC nucleon are present in this section.
The structure function $F_2$ is directly connected to
the parton distribution functions (PDFs).
In the calculations, we take the dynamical PDFs,
which are generated from DGLAP evolution equations \cite{Dokshitzer:1977sg,Gribov:1972ri,Altarelli:1977zs}
with the input of
three valence quark distributions at an extremely low $Q_0^2$.
The initial three valence quark distributions at $Q_0^2$
of the free nucleon are taken from an estimation of the maximum entropy method \cite{Wang:2014lua},
which produce the structure function consistent with the experimental data at high $Q^2$.
In the nucleon swelling model, all the nuclear modifications are reflected
in the increase of the quark confinement space.
Therefore, to evaluate the structure function of the SRC nucleon,
we need just to modify the initial three valence quark distributions
due to the swelling of SRC nucleon.

The enlargement of the confinement size of SRC
can be understood in three different pictures.
(1) In hadron bag model, the high local density reduces the
pressure of the vacuum in which the nucleon embedded,
thus resulting in a bigger size of the nucleon bag.
(2) If the quarks can exchange between the nucleons in the SRC pair,
then it means that the confine space of the quark is increased.
(3) The enlargement of confinement size is also vividly illustrated
with the multiquark cluster model \cite{Jaffe:1982rr,Carlson:1983fs,Chemtob:1983zj,Miller:1984twp,Clark:1985qu}.
When two nucleons form into a six-quark cluster, the confinement space
of this six-quark cluster is naturally larger than the three-quark cluster
(the nucleon) if the quark density is the same.
Moreover, the calculations of Quark-Meson Coupling (QMC) model \cite{Guichon:1987jp,Guichon:1995ue,Saito:2005rv}
and nuclear potential model \cite{Oka:1986vj,Kaelbermann:1994qd,Dukelsky:1995ec}
also give a small deconfinement of the quark in the nuclei.
G. Miller analyzed the elastic electron-nucleus scattering under the Ward-Takahashi identity,
and find that with the input of lattice QCD the off-shell nucleon expands the size \cite{Miller:2019mae}.

There are two ways to apply the nucleon swelling effect to the quark distributions.
(1) A bigger nucleon is equivalent to a higher resolution power
of the photon probe in DIS.
In the language of QCD evolution, the $Q^2$-rescaling
\cite{Close:1983tn,Jaffe:1983zw,Nachtmann:1983py,Close:1984zn,Close:1987ay}
(an higher resolution power) is carried out to interpret the effect.
(2) Due to the change of quark confinement space,
the quark momentum distribution also varies
according to the Heisenberg uncertainty principle.
If the uncertainty of the spatial distribution becomes larger,
the uncertainty of the valence quark distribution reduces accordingly \cite{Wang:2018wfz,Wang:2016mzo}.
The uncertainty of a random variable is quantified with the width of the distribution.
The width can be taken as the standard deviation of the distribution.
Thus the widths of the valence distributions are given by,
\begin{equation}
\begin{aligned}
&\sigma(x_u)=\sqrt{<x_u^2>-<x_u>^2},\\
&\sigma(x_d)=\sqrt{<x_d^2>-<x_d>^2},\\
&<x_u>=\int_0^1 x\frac{u_v(x,Q_0^2)}{2}dx,\\
&<x_d>=\int_0^1 xd_v(x,Q_0^2)dx,\\
&<x_u^2>=\int_0^1 x^2\frac{u_v(x,Q_0^2)}{2}dx,\\
&<x_d^2>=\int_0^1 x^2d_v(x,Q_0^2)dx,
\end{aligned}
\label{eq:WidthDef}
\end{equation}
In this work, we apply the second method to evaluate the PDFs
and the structure function of the SRC nucleon.
We also tried the $Q^2$-rescaling model \cite{Close:1983tn,Jaffe:1983zw}.
The $Q^2$-rescaling model does not generate the anti-shadowing effect,
and the EMC effect produced from the $Q^2$-rescaling model is smaller
than that from our distribution-changing model.

\begin{figure}[htbp]
\begin{center}
\includegraphics[width=0.45\textwidth]{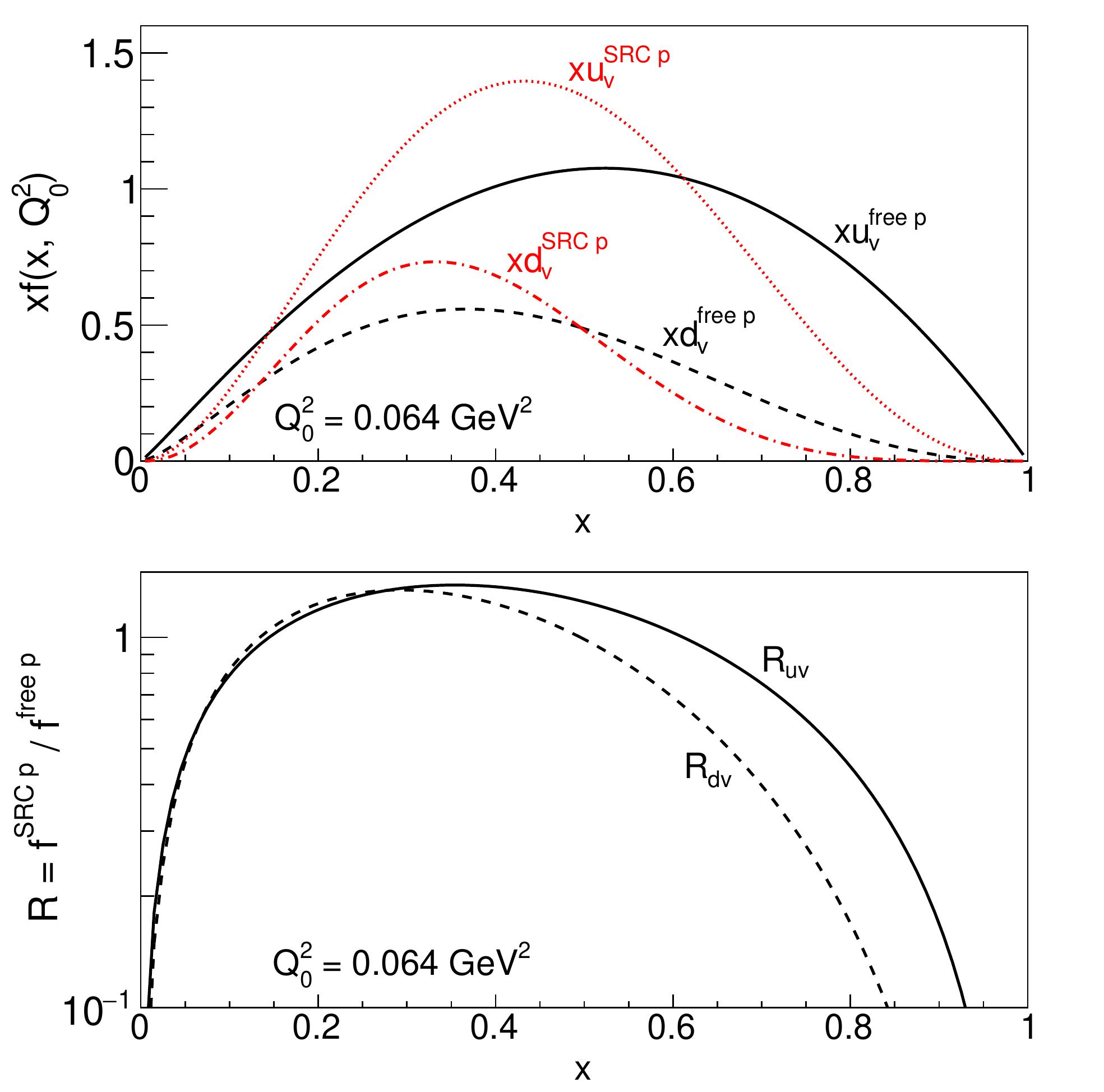}
\caption{
  (color online) The upper panel shows the valence quark distributions of
  the free proton and the SRC proton at the initial scale $Q_0^2$.
  The lower panel shows the nuclear modification ratios of the valence
  quark distributions at the initial scale $Q_0^2$.
  The change of the width of the valence quark distribution in SRC proton is made
  according to the Heisenberg uncertainty principle and the swelling of the quark confinement.
}
\label{fig:ValenceQuarkDistributionsAtQ0}
\end{center}
\end{figure}

The quark confinement space of the six-quark bag from N-N SRC is
twice of that of the nucleon bag, assuming that the quark density is the same.
If we assume the quarks exchange completely freely
between the two nucleons in SRC, the swelling factor of
the quark confinement space also can be as large as two.
We do not have a certain answer for the quark confinement of the SRC pair.
In this work, we try to see the largest nuclear modification
that the SRC nucleons can provide.
If the largest nuclear modification from SRC nucleons can not explain
the EMC effect, then we should look for more origins of the EMC effect.
Therefore, in this work we assume that the quark confinement space
in SRC pair is twice of that in the free nucleon.
Therefore quark confinement radius in SRC pair is then $(2)^{1/3}$
times of that in the free nucleon.
According to the Heisenberg uncertainty principle,
the width of valence quark distribution in SRC nucleon is
reduced by a factor of $(2)^{-1/3}$, which is written as,
\begin{equation}
\begin{aligned}
\frac{\sigma(x_q^{\rm SRC~N})}{\sigma(x_q^{\rm free~N})} = \left(\frac{1}{2}\right)^{1/3},~~(q=u,d).
\end{aligned}
\label{eq:SwellingModel}
\end{equation}
In the calculation, the valence quark distributions of the free nucleon
and the SRC nucleon are all parameterized as the Beta function $Ax^B(1-x)^C$.
The momentum sum rule and the valence sum rule are also required at $Q_0^2$,
which are written as,
\begin{equation}
\begin{aligned}
\int_0^1 x[u_v(x,Q_0^2) + d_v(x,Q_0^2)] dx = 1, \\
\int_0^1 u_v(x,Q_0^2) dx = 2, \\
\int_0^1 d_v(x,Q_0^2) dx = 1.
\end{aligned}
\label{eq:SumRules}
\end{equation}
The benchmark valence quark distributions of the free nucleon are taken from Ref \cite{Wang:2014lua}.
The valence quark distributions of the SRC nucleon are solved with
Eq. (\ref{eq:WidthDef}) and Eq. (\ref{eq:SwellingModel}).
The input valence quark distributions at $Q_0^2$ ($\sim 0.1$ GeV$^2$)
of the free proton and the SRC proton are shown in Fig. \ref{fig:ValenceQuarkDistributionsAtQ0}.
One sees that the nuclear modification at $Q_0^2$
on the valence quark distributions are strong for the SRC nucleon.

With the obtained valence quark distributions at $Q_0^2$,
the PDFs and the structure function at high $Q^2$ are given
by the DGLAP evolution equations \cite{Dokshitzer:1977sg,Gribov:1972ri,Altarelli:1977zs}.
The initial scale $Q_0^2$ and the strong coupling
$\alpha_s$ are taken from Refs. \cite{Wang:2014lua,Wang:2016sfq}.
$Q_0^2$ is set at 0.064 GeV$^2$, where there are only valence quarks at the scale.
The running strong coupling is taken as $\alpha_{\rm s} = 4\pi/[\beta_0{\rm ln}(Q^2/\Lambda^2)]$,
with $\beta_0=11-2n_{\rm f}/3$ and $\Lambda_{\rm LO}^{3,4,5,6}=204,175,132,66.5$ MeV.
The parton-parton recombination correction \cite{Mueller:1985wy,Zhu:1998hg} is included in order to slow
down the fast splitting process due to the large $\alpha_s$ at low $Q^2$.
For the calculations of the neutron PDFs and structure function,
the isospin symmetry of nucleon is assumed, as $u^n = d^p$ and $d^n = u^p$.

\begin{figure}[htbp]
\begin{center}
\includegraphics[width=0.45\textwidth]{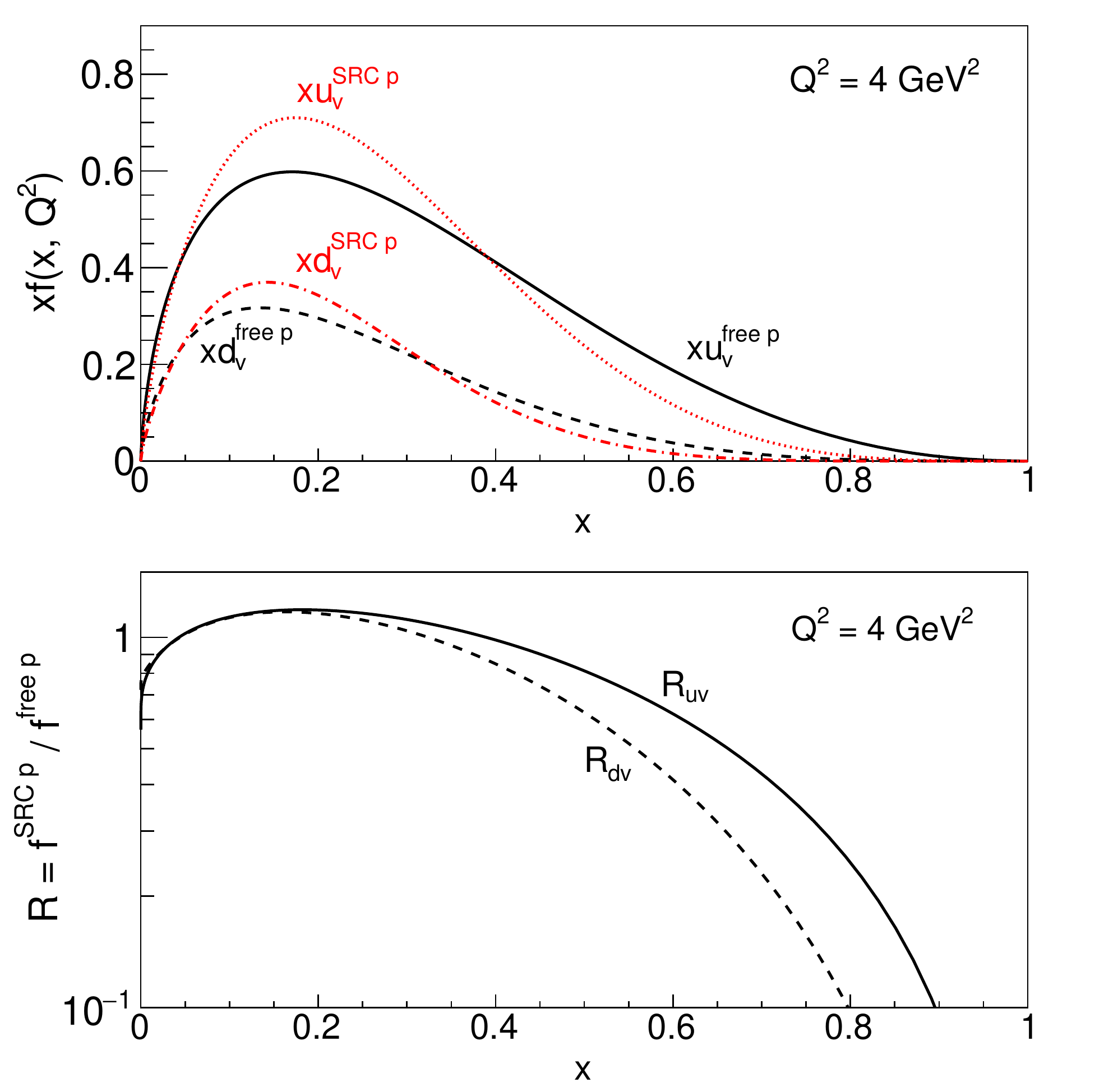}
\caption{
  (color online) The upper panel shows the valence quark distributions of
  the free proton and the SRC proton at the $Q^2$ relevant to the experimental measurements.
  The lower panel shows the nuclear modification ratios of the valence
  quark distributions at the experimental scale $Q^2$.
  The valence quark distributions are given with DGLAP evolution equations
  and the input valence quark distributions at the hadronic scale.
}
\label{fig:ValenceQuarkDistributionsAtQ}
\end{center}
\end{figure}

Applying the DGLAP evolution equations discussed above,
the valence quark distributions at a high $Q^2$ ($4$ GeV$^2$) are
obtained and shown in Fig. \ref{fig:ValenceQuarkDistributionsAtQ}.
The shapes of the valence quark distributions change dramatically
during the evolution from $Q_0^2$ to $Q^2$.
At the high $Q^2$ scale, the valence distributions of
the SRC nucleon are lower than that of the free nucleon
in the intermediate $x$ range of $x\gtrsim 0.35$,
which is consistent with the EMC effect observed in experiment.
The valence-distribution ratios of the SRC nucleon to the free nucleon
are also shown in Fig. \ref{fig:ValenceQuarkDistributionsAtQ} at the high $Q^2$.

\section{Results and discussions}
\label{sec:results}

\begin{figure}[htbp]
\begin{center}
\includegraphics[width=0.42\textwidth]{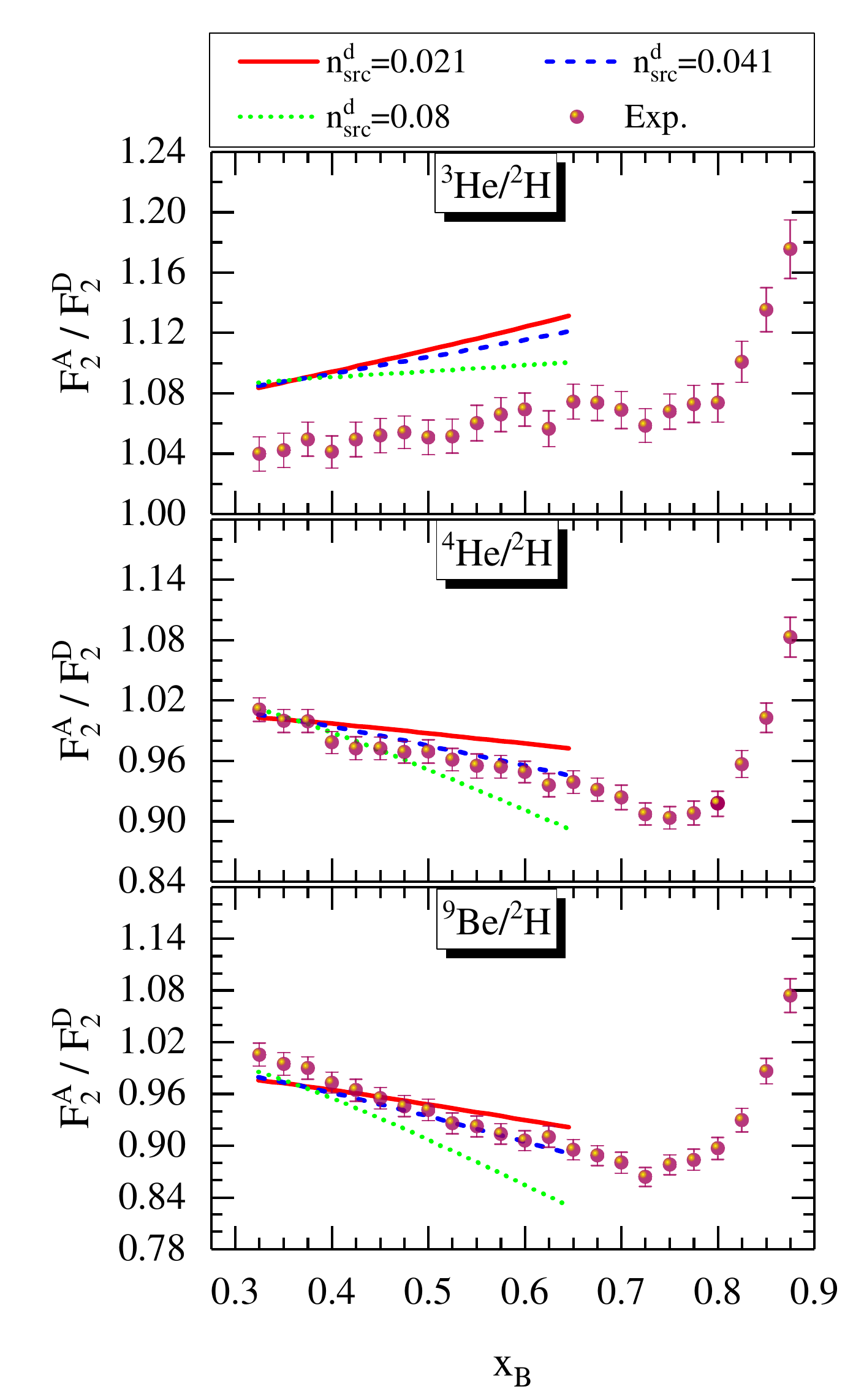}
\caption{
  (color online) Comparisons between our SRC-driven model calculations
  for the EMC effect and the experimental measurements of light nuclei \cite{Seely:2009gt,Arrington:2021vuu}.
  The swelling effect of the SRC nucleon is assumed to be the origin
  of the EMC effect in our calculations.
  The curves of different styles show the results with different input values
  for the parameter $n_{\rm SRC}^{\rm d}$. See the main text for more explanations.
}
\label{fig:LightNucleiEMCEffect}
\end{center}
\end{figure}

\begin{figure*}[htbp]
\begin{center}
\includegraphics[width=0.75\textwidth]{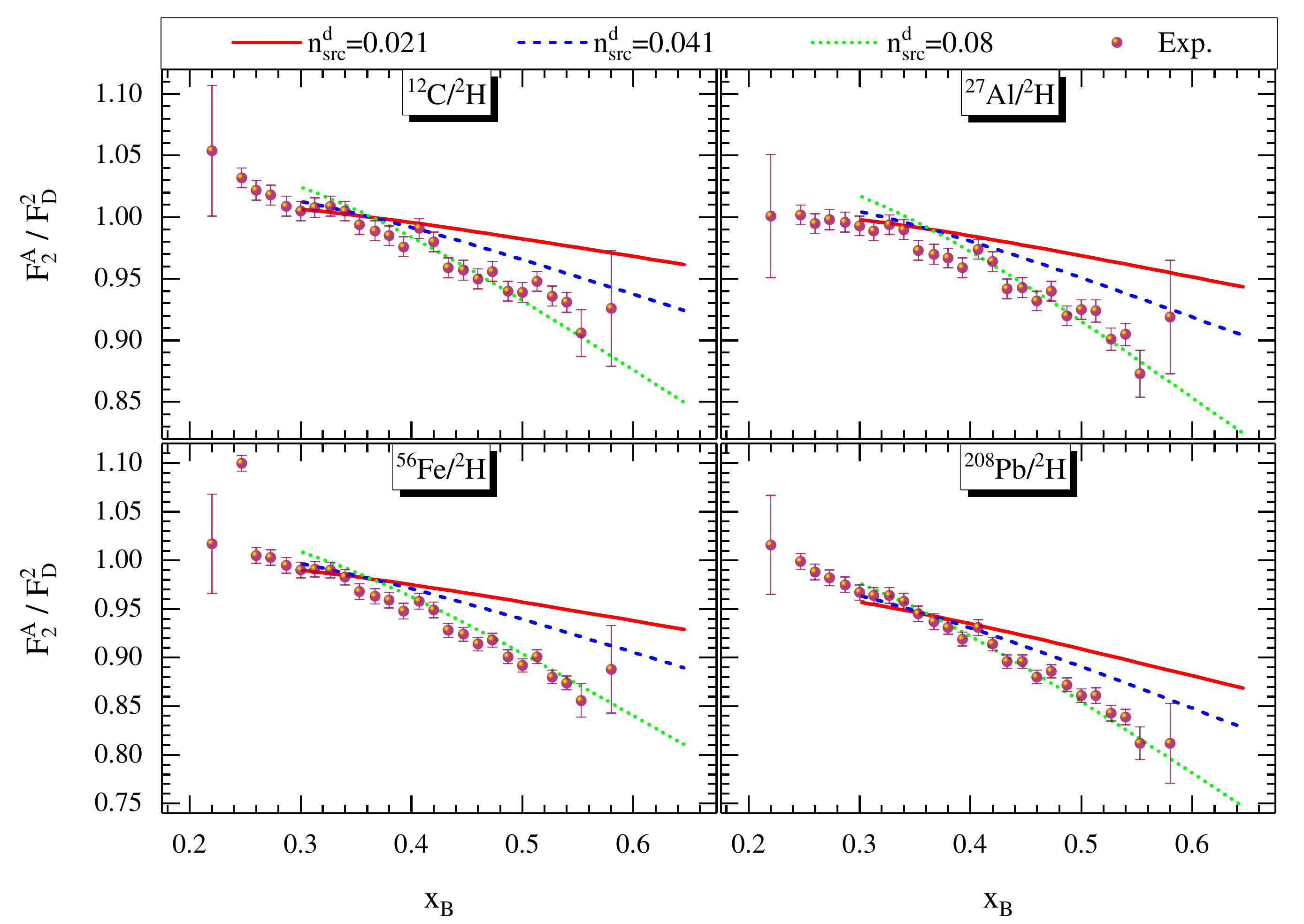}
\caption{
  (color online) Comparisons between our SRC-driven model calculations
  for the EMC effect and the experimental measurements of heavy nuclei \cite{CLAS:2019vsb}.
  The swelling effect of the SRC nucleon is assumed to be the origin
  of the EMC effect in our calculations.
  The curves of different styles show the results with different input values
  for the parameter $n_{\rm SRC}^{\rm d}$. See the main text for more explanations.
}
\label{fig:HeavyNucleiEMCEffect}
\end{center}
\end{figure*}

The predicted EMC ratios based on the assumptions of SRC-driven EMC effect and
the SRC nucleon swelling model are shown in Fig. \ref{fig:LightNucleiEMCEffect}
and Fig. \ref{fig:HeavyNucleiEMCEffect}, for light nuclei and heavy nuclei respectively.
The experimental data are taken from the analyses by CLAS collaboration \cite{CLAS:2019vsb}
and JLab Hall C collaboration \cite{Seely:2009gt,Arrington:2021vuu}.
The number of SRC pairs in deuteron are estimated to be from 0.021 to 0.041.
$n_{\rm SRC}^{\rm d} = 0.021$ is obtained from the fit to the correlation between
the nuclear mass and the SRC scaling ratio $a_2$ \cite{Wang:2020egq}.
$n_{\rm SRC}^{\rm d} = 0.041$ is estimated by counting the nucleons of momentum
above $k_F\approx 275$ MeV/c \cite{CLAS:2005ola,CLAS:2018yvt}. For light nuclei, one sees that
the EMC effect from SRC nucleons can reproduce the experimental data within
our nucleon swelling model and with $n_{\rm SRC}^{\rm d} = 0.041$.
However, for the heavy nuclei, our model calculations from the swelling SRC nucleons
are not enough to explain the experimental observations, with either
$n_{\rm SRC}^{\rm d} = 0.021$ or $n_{\rm SRC}^{\rm d} = 0.041$.

\begin{figure}[htbp]
\begin{center}
\includegraphics[width=0.42\textwidth]{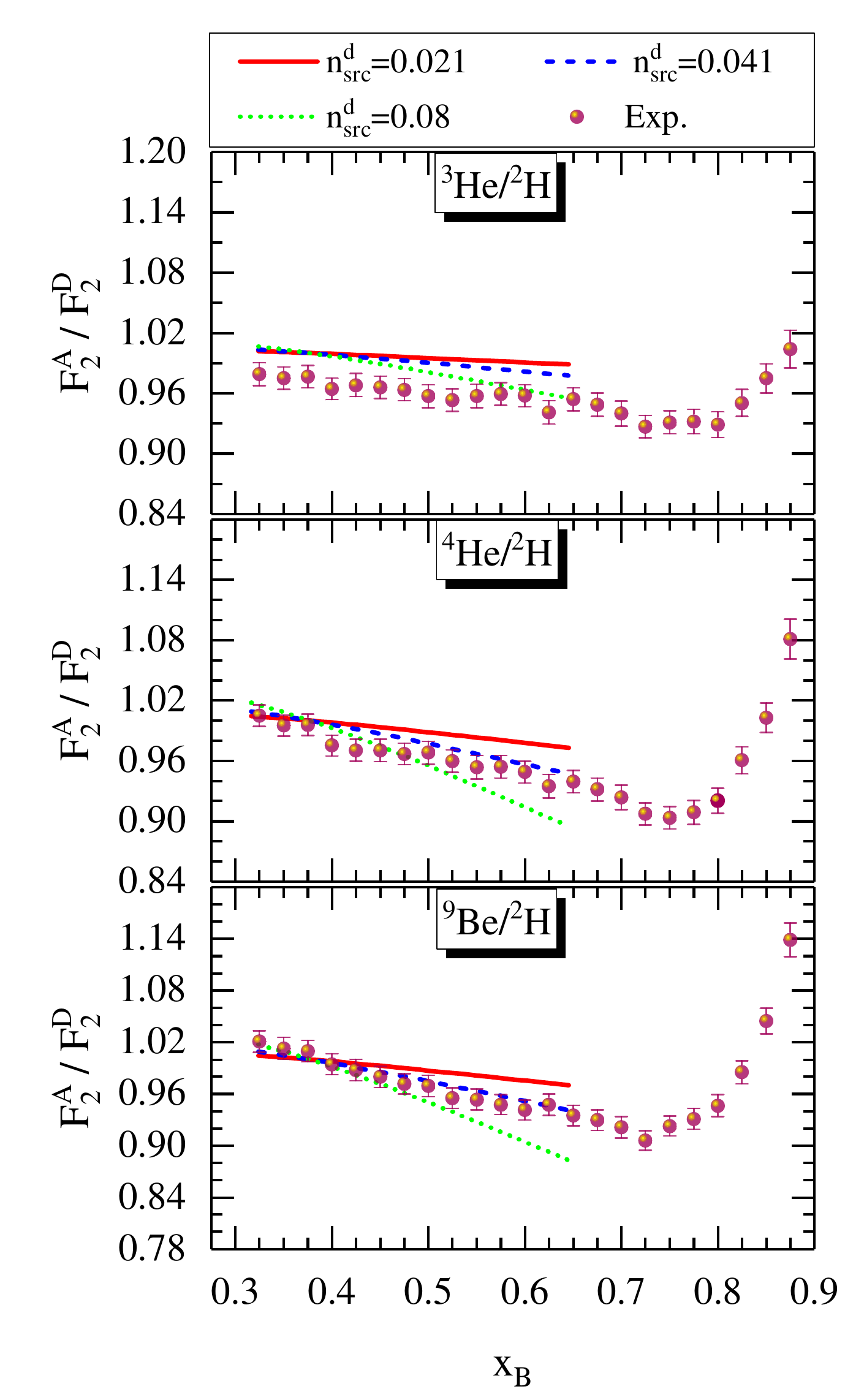}
\caption{
  (color online) Comparisons between our SRC-driven model calculations
  for the EMC effect and the experimental measurements of light nuclei \cite{Seely:2009gt,Arrington:2021vuu},
  with the isoscalar corrections for both the theoretical calculations and the experimental data.
  The swelling effect of the SRC nucleon is assumed to be the origin
  of the EMC effect in our calculations.
  The curves of different styles show the results with different input values
  for the parameter $n_{\rm SRC}^{\rm d}$. See the main text for more explanations.
}
\label{fig:LightNucleiEMCEffect_isoscalar}
\end{center}
\end{figure}

In Fig. \ref{fig:LightNucleiEMCEffect}, the discrepancy between our model
and the experimental data is big for $^3$He, which hints that the $d(x)/u(x)$
ratio may not be consistent with the real $F_2^{\rm n}/F_2^{\rm p}$ measurements.
To minimize the influence of the un-tuned $d(x)/u(x)$ ratio in our model,
we also made the comparisons for the isoscalar corrected EMC effect,
which is shown in Fig. \ref{fig:LightNucleiEMCEffect_isoscalar}.
The experimental data of the EMC effect with isoscalar corrections are taken
from Ref. \cite{Arrington:2021vuu}. The proton number and the neutron number are
required to be the same in the theoretical calculations accordingly.
One finds that the big disagreement between our model and the data is reduced for $^3$He.
And the conclusion does not change for the isoscalar corrected EMC effect.
For light nuclei, the EMC effect merely from SRC nucleons can reproduce
the experimental data within our nucleon swelling model
with $n_{\rm SRC}^{\rm d} = 0.041$.

In order to explain the EMC effect of heavy nuclei,
the parameter $n_{\rm SRC}^{\rm d}$ in our model should be tuned up to 0.08.
However, with $n_{\rm SRC}^{\rm d}=0.08$ our model can not reproduce the EMC effect of light nuclei.
More importantly, $n_{\rm SRC}^{\rm d}=0.08$ is not consistent with
the previous estimations by counting the high-momentum nucleons above Fermi motion region \cite{CLAS:2005ola,CLAS:2018yvt}.
In order to explain the contradiction, we speculate that the universality
of SRC nucleon structure is violated, or there are more origins of the EMC effect
for the heavy nuclei in order to agree with the experimental observations.
And other origins for the EMC effect have nuclear dependence
from light nuclei to heavy nuclei.
A previous analysis also suggested that the underlying physics of the
EMC effect for the heavy nuclei is different from that for the light nuclei \cite{GarciaCanal:2012adh}.

\begin{figure*}[htbp]
\begin{center}
\includegraphics[width=0.75\textwidth]{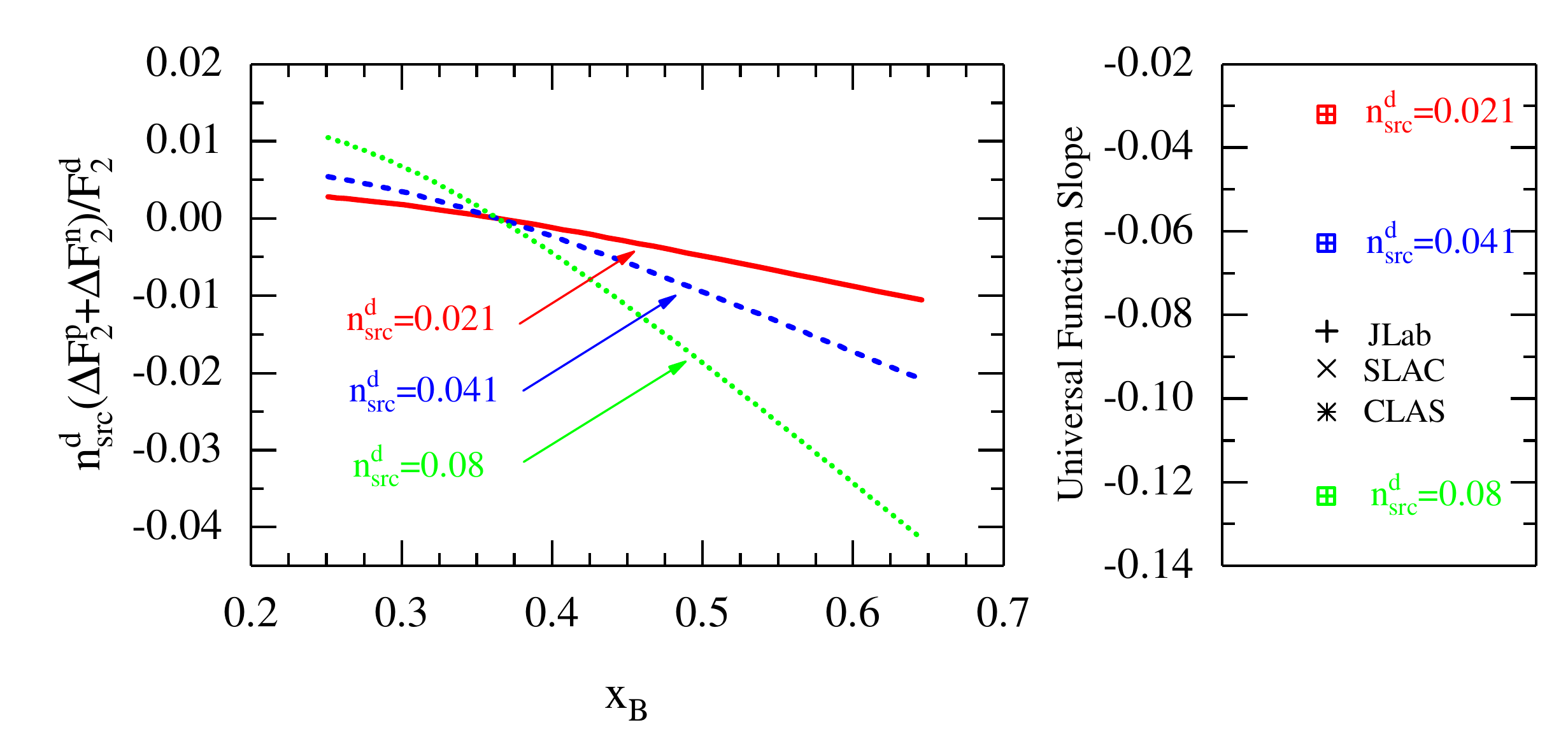}
\caption{
  The universal modification function for the structure function
  of the SRC nucleon inside the deuteron
  calculated in the nucleon swelling model.
  The curves of different styles show the results with different
  input values for the parameter $n_{\rm SRC}^{\rm d}$.
  In the right panel, the slopes of modification functions are shown.
  The extracted slopes of the universal function are taken from Ref. \cite{CLAS:2019vsb},
  and the original experimental data are from SLAC \cite{Gomez:1993ri},
  JLab Hall C \cite{Seely:2009gt,Arrington:2021vuu}, and CLAS \cite{CLAS:2019vsb}.
  See the main text for more explanations.
}
\label{fig:EMCUniversalFunction}
\end{center}
\end{figure*}

The universal modification function of the SRC nucleon in the deuteron is calculated
and shown in Fig. \ref{fig:EMCUniversalFunction},
based on the nucleon swelling model discussed in the previous section.
The slope of the universal modification function is also evaluated
by CLAS collaboration from the experimental data at SLAC \cite{Gomez:1993ri},
JLab \cite{Seely:2009gt,Arrington:2021vuu}, and CLAS \cite{CLAS:2019vsb},
which are shown in Fig. \ref{fig:EMCUniversalFunction}.
The experimental extractions give the slope in a range from about 0.08
to 0.11, consistently.
Our model predictions are weaker than the result from the experimental analysis
in terms of the slope of the universal modification function,
with $n_{\rm SRC}^{\rm d} = 0.021$ and $n_{\rm SRC}^{\rm d} = 0.041$.
Therefore, one may conclude that either the assumption
that the EMC effect only comes SRC nucleons is wrong,
or the universality of SRC nucleon structure is violated,
or the nucleon swelling model for SRC nucleon needs improvement.

With the recent analysis of the experimental data from JLab Hall C \cite{Arrington:2021vuu},
the EMC effects for the heavy nuclei are found to be weaker than
those measured by CLAS collaboration \cite{CLAS:2019vsb}.
Therefore there are also small inconsistences among the experiments.
This kind of small inconsistence is also shown in the slopes
of the universal modification function in Fig. \ref{fig:EMCUniversalFunction}.
Anyway, these differences among different experiments can be evaluated
and removed with more experiments, improved apparatus and analysis method.

In our model calculations for the SRC-induced EMC effect,
the pair c.m. motion effect is not considered.
In a heavy nucleus, this effect reduces about 20\% of the probability
of a nucleon being in the SRC correlation \cite{CLAS:2005ola,Fomin:2011ng}.
Thus, considering the pair c.m. motion effect, the predicted
EMC effect in our model also decreases about 20\% for the heavy nucleus.
The discrepancy between the model prediction and the experimental data
however increases slightly for the heavy nuclei.
Therefore the conclusions given in this work do not change,
with the consideration of the pair c.m. motion effect. 
In our model, the 20\% reduction in the SRC scaling factor 
of the heavy nucleus corresponds to a 20\% increase in the parameter $n_{\rm SRC}^{\rm d}$, 
to reproduce the same magnitude of the EMC effect.

\section{Summary}
\label{sec:summary}

We have tested the hypothesis that the N-N SRC is the dominant source for the nuclear EMC effect.
Based on the nucleon swelling model for the SRC nucleon
and that the number of SRC pairs in deuteron is about 0.041,
we find that the nuclear corrections on the SRC nucleons more or less explain
the nuclear EMC effect of the light nuclei.
However, with the same model and inputs, only the nuclear modifications on the
SRC nucleons can not reproduce the nuclear EMC effect of the heavy nuclei.
We guess that the inner structure of the mean-field nucleon is also modified,
or the SRC universality is violated, or there are more origins for the EMC effect
beyond the N-N SRC. Although the SRC universality is favored in experiments,
our analysis hints that the modification on the structure function of SRC nucleon
may be stronger in the heavy nuclei compared to that of the light nuclei.
Another explanation is that there are more origins for the EMC effect
(such as 3N and 4N SRCs) and the number of these multi-nucleon SRC pairs does not
linearly scale with the number of N-N SRC pairs.

Based on the current knowledge of the number of p-n SRC pairs in deuteron
and the nucleon swelling model for the modification of valence quark distributions,
our obtained universal modification function of the SRC nucleon
$n_{\rm SRC}^{\rm d} (\Delta F_2^{\rm p} + \Delta F_2^{\rm n}) / F_2^{\rm d}$ is
not consistent with the analysis of the experimental data.
The experimental extraction of universal modification function of SRC nucleon is
performed with the assumption that the EMC effect is completely driven by N-N SRC.
Based on the analysis in this work, we conclude that there is the correlation between
the N-N SRC strength and the EMC effect, but there is not a causal relation
between these two phenomena.
This conclusion is consistent with the recent results
from the calculations of the off-shellness correction \cite{Wang:2020uhj}
and the $x$-rescaling model \cite{Wang:2022kwg} for the SRC nucleon.

\begin{acknowledgments}
This work is supported by the National Natural Science Foundation of China under the Grant NO. 12005266
and the Strategic Priority Research Program of Chinese Academy of Sciences under the Grant NO. XDB34030301.
N.-N. Ma is supported by the National Natural Science Foundation of China under the Grant NO. 12105128.
\end{acknowledgments}

\bibliographystyle{apsrev4-1}
\bibliography{refs}

\end{document}